\documentclass[3p,times,twocolumn]{elsarticle}

\usepackage{ecrc}

\volume{00}

\firstpage{1}

\journalname{Nuclear Physics B Proceedings Supplement}

\runauth{D.A.~Dwyer and L.~Ludhova}

\jid{nuphbp}

\jnltitlelogo{Nuclear Physics B Proceedings Supplement}

\usepackage{amssymb}
\usepackage[figuresright]{rotating}

\newcommand{\thet}[1]{$\theta_{#1}$}
\newcommand{\dmSq}[1]{$\left|{\Delta}m^2_{#1}\right|$}
\newcommand{\nubar}[1]{$\overline{\nu}_{#1}$}

\usepackage{color}
\newcommand{\modf}[1]{\textcolor{blue}{#1}}
\definecolor{cmntcol}{rgb}{0,0.5,0}

\usepackage[normalem]{ulem}

\begin{document}

\begin{frontmatter}



\dochead{}

\title{Oscillations at low energies}


\author[LBNL]{D.A.~Dwyer}
\author[INFN]{L.~Ludhova}

\address[LBNL]{Lawrence Berkeley National Laboratory, Berkeley, CA 94720, USA}
\address[INFN] {Istituto Nazionale di Fisica Nucleare, Milano, 20133, Italy}

\begin{abstract}
A concise summary of the ``Oscillation at low energies''
 parallel
session at the 2014 Neutrino Oscillation Workshop is
 provided.
Plans to use man-made neutrinos and antineutrinos to
 determine the
neutrino mass hierarchy, search for sterile
 neutrinos, and to
observe coherent neutrino-nucleus scattering were
 discussed.
Potential measurements of solar neutrinos, supernova
 neutrinos, and
geoneutrinos are also summarized.

\end{abstract}


\end{frontmatter}

%\linenumbers

\section{Introduction}
\label{intro}

Measurements of the oscillation of reactor, solar, accelerator,
  and atmospheric neutrinos have given us a fairly consistent picture
  of the flavor mixing of massive neutrinos~\cite{PDG2014}.  The
  topics covered during the ``Oscillation at low energies'' parallel
  session of the Neutrino Oscillation Workshop focused on the
  questions addressed by future measurements using low energy
  neutrinos.  Plans are proceeding toward determination of the
  neutrino mass hierarchy using medium-distance (60~km) oscillation of
  reactor antineutrinos.  Multiple searches for oscillation to sterile
  states at short-distances ($\sim$10~m) from reactor or intense
  radioactive sources are expected to occur over the next few
  years. To constrain models of the emission of antineutrinos from
  nuclear reactors, theorists request additional measurements.
  Advancements in low-energy-threshold detectors allow for potential
  measurement of coherent neutrino-nucleus scattering.
Solar neutrino measurements are an important ingredient for solar
models and for testing the LMA-MSW solution for neutrino
oscillations. Supernova neutrinos detection would provide a lot of
information about star collapse and neutrino physics, and
thus several running and future experiments are focused toward this goal. The effects of decoherence by wave
packet separation on complex collective phenomena which could manifest
in neutrino oscillations in the extreme supernovae environment have
recently been studied. Neutrino geoscience is a new
interdisciplinary field aiming to measure the Earth's radiogenic heat, providing an important benchmark for
geosciences.
These topics are briefly summarized here.

\section{Man-made sources of $\nu$ and $\overline{\nu}$}
\label{manmade}

\subsection{Neutrino mass hierarchy}
\label{hierarchy}

% JUNO: M. He
Due to the non-zero value of \thet{13}, a signature of the neutrino
mass hierarchy is present in the oscillation of antineutrinos emitted
by nuclear reactors~\cite{Petcov2002, Learned2006}.  The neutrino
mass-squared differences \dmSq{ji}= $m^2_j - m^2_i$ cause
energy-dependent \nubar{e} disappearance, introducing a distortion of
the reactor \nubar{e} energy spectrum.  The oscillation frequencies
caused by \dmSq{31} and \dmSq{32} differ by only 3\% according to
measurements of \dmSq{21}.  Measurements of the mixing angles predict
that the \nubar{e} oscillation amplitude from \dmSq{31} is twice as
large as that from \dmSq{32}.  The mass hierarchy can be discerned by
whether the smaller-amplitude \dmSq{32} oscillation is at a slightly
lower (normal) or higher (inverted) frequency.  Discrimination of
these two frequencies from the distortion of the reactor \nubar{e}
energy spectrum is most pronounced at the first \dmSq{21} oscillation
maximum, or roughly $\sim$60~km from a reactor.

Two experiments have been proposed to determine the mass hierarchy
using reactor \nubar{e}: the Jiangmen Underground Neutrino Observatory
(JUNO)~\cite{Wang2014} and the RENO-50 experiment~\cite{Kim2013}.  The
JUNO experiment will be located in an new underground laboratory
(700~m rock overburden) in Kaiping city, Jiangmen region, Guangdong
province in southern China.  Reactor antineutrinos will be produced by
the Yangjiang and Taishan commercial reactor facilities, each located
53~km from the planned site.  Each facility will have six reactors,
with a total power of 17.4~GW$_{\rm th}$ and 18.4~GW$_{\rm th}$
respectively.  Both facilities are currently under construction, with
26.6~GW$_{\rm th}$ of power in operation by 2020.

As designed, the JUNO detector consists of a 20~kton target of liquid
scintillator contained in a 35.4~m-diameter transparent acrylic
sphere.  Light produced by \nubar{e} interactions in the target is
detected by 15000 20-inch photomultiplier tubes (PMTs) mounted on a
stainless steel frame surrounding the target.  The entire system is
contained within an active water Cherenkov veto detector, instrumented
with 1500 additional PMTs.  A secondary planar veto detector rests
above the Cherenkov detector.  Approximately $\sim$40 reactor
\nubar{e} inverse beta decay interactions per day are predicted within
the target volume.  Preliminary estimates for background from
accidentals, fast neutrons, and long-lived spallation isotopes are
$\sim$10\%, 0.4\%, and 0.8\% respectively.

According to Ref.~\cite{Li2013}, after six years of operation the JUNO
experiment can discriminate between the two hierarchies with a
$\Delta\chi^2 > 9$.  By 2020, existing experiments may measure
\dmSq{32} with 1.5\% precision.  Including this external constraint,
the JUNO sensitivity is predicted to improve to $\Delta\chi^2 > 16$.
The sensitivity of the experiment is strongly dependent on the energy
resolution of the detector.  Statistical fluctuation in the number of
detected photons is predicted to limit the resolution.  To achieve the
desired energy resolution of 3\%/$\sqrt{E {\rm [MeV]}}$, the JUNO
detector must collect an average of five times more photons than Daya
Bay detectors.  Current research and development is focused on
increasing the collected light by scintillator purification to reduce
light attenuation, exploring new designs for high efficiency PMTs, and
maximizing the PMT surface coverage.  The mechanical integrity,
stresses, and aging of the acrylic target container are also under
study.  To reduce the cabling required for the large number of PMTs,
the collaboration is testing prototype electronics for underwater
installation close to the PMTs.  Aside from the mass hierarchy, the
JUNO experiment intends to measure \thet{12}, \dmSq{21}, and \dmSq{31}
to better than 1\% precision.  Measurements of solar neutrinos,
atmospheric neutrinos, geoneutrinos, sterile neutrinos, proton decay,
and other exotic models are being explored.

The JUNO experiment has obtained the land for the detector site and
completed a geologic survey.  The civil design is almost complete, and
construction is expected to begin soon and complete in 2017.  Detector
construction is planned for 2017 through 2019.  Filling of the
detector with liquids and collection of the first data is planned for
2020.

\subsection{Sterile neutrinos}
\label{sterile}

 % Anomaly: P. Huber

Measurements of the the emission of antineutrinos from nuclear
reactors show a deficit relative to recent
estimates~\cite{Mention2011}.  A similar deficit has been observed in
geochemical measurements of neutrinos from intense radioactive
sources~\cite{Abdur2006}. Combined, these discrepancies have been
interpreted as possible evidence of oscillation to hypothetical
sterile neutrino states~\cite{Abaza2012}.

On average, six \nubar{e} are emitted by the beta decays of the
fragment isotopes from a single fission.
%Approximately two of these
%decays emit \nubar{e} with energy above the 1.8~MeV energy threshold
%for detection via inverse beta decay.  
Direct calculation of reactor \nubar{e} emission is complicated by the
lack of detailed knowledge of the energy levels and decay spectra for
the $\sim$10000 unique beta decays which contribute.  Instead,
existing models rely on prediction of the cumulative \nubar{e} flux
using measurements of the corresponding cumulative $\beta^-$ emitted
by fissioning isotopes~\cite{Mueller2011,Huber2011}.  These
predictions account for weak magnetism, radiative, finite nuclear
size, and screening corrections which impact the correspondence
between the $\beta^-$ and \nubar{e} energy spectra.  Recent
reevaluation increased the predicted \nubar{e} flux by 6\% relative to
previous estimates.  Half of the increase is a result of improved
modeling of the correlation between the cumulative energy spectra of
$\beta^-$ and \nubar{e} from fission fragments.  The remaining
increase is caused by inclusion of long-lived fission daughter
isotopes, and updated measurements of the neutron lifetime which
increase the predicted cross section for \nubar{e} interaction with
the detector.  Measurements of the reactor \nubar{e} flux are on
average 6\% lower than these recent predictions.

Preliminary measurements of the energy spectra of reactor \nubar{e}
reported this past summer disagree with the prevailing
models~\cite{RenoNu2014, Abe2014, DybICHEP2014}.  In particular, the
data show a $\sim$10\% excess in the region of 5 to 7 MeV relative to
prediction.  An alternate model relying on direct calculation using
the ENSDF nuclear database predicts a spectrum consistent with these
measurements, although uncertainties in this calculation are
significant~\cite{Dwyer2014}.  Direct calculations of the
corresponding $\beta^-$ spectra also disagree with the measurements
from Refs.~\cite{Feili1985, Schre1985, Hahn1985}.  The measured
\nubar{e} and cumulative $\beta^-$ spectra seem inconsistent, although
including spectral corrections from forbidden beta decays may be able
to accommodate both.

% STEREO, NUCIFER: M. Pequignot 
Multiple experiments are searching for evidence of sterile neutrino
oscillation at low energies.  The NUCIFER experiment attempted an
initial measurement of the total flux of \nubar{e} at 7~m from the
70~MW$_{\rm th}$ Osiris reactor in Saclay, France~\cite{Port2010}.
Light from nubar{e} inverse beta decay interactions with a 850~liter
target of liquid scintillator was detected using 16 PMTs.  The
scintillator was loaded with 0.2\% Gadolinium in order to identify
neutrons by the resulting $\sim$8~MeV of gamma rays emitted following
neutron capture on these nuclei.  Pulse shape discrimination improved
identification of \nubar{e} interactions from background due to
energetic neutrons.  With a detection efficiency of $\sim$30\%, a rate
of $305\pm29$ \nubar{e} interactions per day was measured from April
to May 2013.  The sensitivity was limited by a large background of
high energy gamma rays in close proximity to the reactor.

The Stereo experiment is a successor to the Nucifer experiment.  It
will search for oscillation to sterile states using \nubar{e} emitted
by the 58~MW$_{\rm th}$ compact highly-enriched uranium core at the
Institut Laue-Langevin (ILL) in Grenoble, France.  The detector is
sensitive to oscillation via the relative rates and energy spectra
measured by six 0.3~m$^3$ cells of Gd-loaded liquid scintillator
arranged along the direction of \nubar{e} emission.  Four PMTs collect
the light produced by interactions in each cell.  Based on Geant4
simulations, the detector will have 11\% energy resolution for 2~MeV
positrons, and have $\sim$60\% efficiency for neutron detection.
Validation of a prototype detector is planned for Sep.~2014.
Extensive gamma ray and neutron shielding (lead, polyethylene, B$_4$C)
will be used to mitigate backgrounds from the research neutron beam
lines at ILL.  Two additional scintillator cells, one at the front and
one at back of the detector, will veto external background and improve
containment of gamma rays generated in the target.  An signal rate of
400 \nubar{e} per day is expected.  With 300 days of data, sensitivity
to 5\% sterile mixing with \dmSq{41} of 2~eV$^2$ at the 95\%
confidence level is predicted.  The experiment plans to begin
operation in Spring~2015.

% PROSPECT: T. Langford
The PROSPECT experiment proposes to search for disappearance of
reactor \nubar{e} from the High Flux Isotope Reactor (HFIR) at Oak
Ridge National Laboratory~\cite{Ashe2013}.  This reactor uses a
compact ($\sim$0.5~m diameter) 85~MW highly-enriched uranium core.
The first phase of the experiment consists of a 2.5~ton
optically-segmented liquid scintillator detector located at $\sim$7~m
from the reactor core.  The estimated signal rate is
$\sim$1000~\nubar{e} inverse beta decay interactions per day within
this detector.  Using the variation in the measured signal between
detector segments over one year, 5\% sterile mixing with
\dmSq{41}$\simeq$2~eV$^2$ can be tested at the 3$\sigma$ confidence
level.  The second phase adds an additional 20~ton detector at
$\sim$18~m from the reactor, with 5$\sigma$ sensitivity over the
region suggested by existing anomalies.  The \nubar{e} energy spectrum
from the HFIR reactor is dominated by fission of $^{235}$U, and
measurement would provide a stringent test of \nubar{e} emission
models.  Neutron and gamma ray detectors have been used to make recent
measurements of the background radiation at the HFIR site.
Significant spatial and temporal variations in the backgrounds have
been found.  Observed gamma rays with energies up to 10~MeV are
attributed to thermal neutron capture on materials in the vicinity of
the reactor.  Lithium-doped liquid scintillator has been developed to
improve localization of the \nubar{e} interaction position via n +
$^6$Li $\rightarrow$ {$\alpha$} + $^3$H.  Initial tests using a $^{252}$Cf
radioactive source also show significant pulse-shape discrimination of
neutron captures on $^6$Li from gamma ray interactions, suggesting
high-efficiency background rejection for the reactor \nubar{e}
measurement.

% SOX: J. Gaffiot
Instead of using reactor \nubar{e}, the SOX experiment will use
\nubar{e} and $\nu_e$ emitted by intense radioactive sources to test
sterile neutrino models~\cite{Bell2013}.  The existing Borexino detector at
the Laboratori Nazionali del Gran Sasso (LNGS) provides a convenient
instrument to measure the neutrinos emitted by radioactive
sources. The detector consists of a 300~ton spherical target of liquid
scintillator suspended within a transparent nylon vessel.  The target
is surrounded by a 2.4~m thick transparent layer of buffer liquid
shielding it from external radioactivity.  Scintillation light is
detected by 2212 PMTs.  This inner detector is contained within a
2.1~kton water cherenkov veto detector to reject cosmic ray induced
backgrounds.  Intense radioactive sources of \nubar{e} and $\nu_e$
will be placed in a cavity 8.5~m below the center of the detector.
The initial measurement will use a $^{144}$Ce-$^{144}$Pr antineutrino
source.  The slow beta decay of $^{144}$Ce gives the source a
relatively long half-life of 285~days.  The subsequent fast beta decay
of $^{144}$Pr emits \nubar{e} with a maximum energy of 3.0~MeV, above
the 1.8~MeV threshold for detection via inverse beta decay.  A 100~kCi
source is currently being manufactured by reprocessing spent nuclear
fuel at the Mayak facility in Russia.  Measurement is planned to begin
at the end of 2015, and is predicted to be sensitive to 5\% sterile
mixing at the 95\% confidence level after a year of operation.  A
$^{51}$Cr source emits monoenergetic 0.75~MeV neutrinos via electron
capture with a half-life of 27.7~days.  These neutrinos can be
detected via elastic scattering on $e^-$ in the scintillator target.
Capsules containing a total of 36~kg of chromium, enriched to 38\% of
$^{50}$Cr, will be neutron-activated to $\sim$5~MCi at the High Flux
Isotope Reactor (HFIR) at Oak Ridge National Laboratory.  Given the
lower cross-section and larger background rate for $\nu_e$ detection,
the source will need to be activated and measured twice to reach
sensitivity similar to the $^{144}$Ce experiment.  Transport of these
intense sources from the production sites to LNGS requires significant
planning.

\subsection{Coherent neutrino scattering}
\label{coherent}

% CENNS: P. Barbeau
Coherent scattering of neutrinos on nuclei (CENNS) is a yet unobserved
weak interaction process~\cite{Free1974}.  The coherent recoil of all
nucleons in the nucleus enhances the cross-section as $N^2$.  The
resulting signal is a low energy nuclear recoil, visible as
$\sim$1~keV electron-equivalent ionization in a detector.  Measurement
of CENNS could be relevant to a wide range of topics including
supernova dynamics~\cite{Wils1974}, sterile neutrinos~\cite{Druk1984},
and precision tests of the weak interaction~\cite{Barr2005,Scho2006}.
A collaboration has recently been formed to organize disparate
detector development toward a coordinated measurement of CENNS using
the Spallation Neutron Source (SNS) at Oak Ridge National Laboratory.
The decay-at-rest of spallation pions and muons provide a flux of
$2\times10^7$ neutrinos cm$^{-2}$ s$^{-1}$ at a distance of 20~m from
the SNS.  Backgrounds uncorrelated with the SNS pulses are negligible
due to the low duty factor $\sim$$4\times10^{-5}$.  Background from
energetic neutrons produced by the SNS are significant.  Measurements
of the neutron flux using a coded aperture neutron detector array
identified the SNS basement as promising detector location at 20~m
from the source.  Additional background is predicted from neutrons
produced by neutrino inelastic excitation of lead commonly used for
detector shielding.  A in-situ measurement of this background is
imminent, with an expected detection of $\sim$300 neutrons over 60
days of operation.  Measurement of this process is relevant for the
HALO supernova neutrino detector~\cite{Shan2010}, as well as for
supernova dynamics~\cite{Qian1997,Mcla2004}.  Measurement of
neutrino-inelastic neutron emission for other elements (Fe, W, Cu, Bi,
etc.) is also being considered.

 \section{Natural sources of $\nu$ and $\overline{\nu}$}
\label{natural}

\subsection{Solar neutrinos}
\label{solar}

  % Theory: F. Villante
  % Super-K Matter Eff.: M. Smy
  % 7-Be Detection: Y. Takemoto
  % SNO+: F. Descamps

The theoretical solar neutrino fluxes are predicted by the Standard Solar Model (SSM)~\cite{SSM2011}, based on the stellar structure equations,  solar chemical evolution paradigm, and observational constraints, including the surface composition.  A reevaluation of the latter (AGSS09~\cite{LOWZ}) led to substantially lower abundances of heavy elements ($Z > 2$) with respect to the previous work GS98~\cite{HIGHZ}.  The solar metallicity problem arose: the new AGSS09 compositional model, to the contrary of the older GS98,  leads to a strong discrepancy between helioseismological data and the SSM predictions. No fully satisfactory solution to this puzzle has been found yet, including attempts to vary solar opacity or to consider non-standard chemical evolution of the Sun. Since the predicted solar-neutrino fluxes depend on the solar metallicity, their measurement can help in solving this puzzle. Recently, the abundances of volatiles (C, N, O, Ne)  and refractories  (Mg, Si, S, Fe) has been inferred from helioseismic and solar neutrino data~\cite{Villante}. The best-fit abundances are consistent at 1$\sigma$ with GS98, however it is not possible to disentangle the degeneracies among individual elements.

The primary reaction of the so called $pp$-cycle, a sequence of nuclear reactions that converts hydrogen into helium and powers the Sun, is  the fusion of two protons with the emission of a low-energy neutrino. These so-called $pp$ neutrinos constitute nearly the entirety of the solar neutrino flux. Borexino collaboration has reported the first spectral observation of these neutrinos~\cite{ppborex} with 11\% precision, demonstrating that about 99\% of the power of the Sun, 3.8431033 ergs per second, is generated by the proton-proton fusion process. This measurement also confirms the solar stability at $10^5$-years time scale, by showing that at this level of precision the photon  and neutrino luminosities are in agreement.

The GS98 and AGSS09 models predict the flux of  0.862\,keV $^7$Be solar neutrinos which differs by about  9\%~\cite{SSM2011, Villante}. The best value of the 5\% precision measurement of the interaction rate of these neutrinos, achieved by Borexino~\cite{be7BX}, falls just in between the two predictions; thus, new measurements are still important.  After an extensive purification of the 1\,kton of liquid scintillator in years 2008-2009, the KamLAND  collaboration has performed the second direct measurement of the $^7$Be neutrinos~\cite{be7KL} with 15\% precision. Utilizing a global three flavour oscillation analysis, the observed rate of $582 \pm 90$ (kton $\cdot$ day)$^{-1}$ corresponds to the $^7$Be solar neutrino flux of $(5.82 \pm 0.98) \times 10^9$ cm$^{-2}$ s$^{-1}$, which is consistent with both SSM predictions. 

The main goal of the SNO${^+}$ experiment~\cite{SNO+}, placed in the SNOLAB in Canada, is the measurement of the neutrino-less double-beta decay ($0\nu-\beta\beta$) by means of 0.3\% loading of scintillator with 800 kg of $^{130}$Te (using natural Te) planned for spring 2016.  The detector will be filled with water in spring 2015, while with 1\,kton of LAB scintillator in fall 2015. $^8$B solar neutrinos are one of the irreducible backgrounds for the $0\nu-\beta\beta$ measurement and  are planned to be measured by SNO$^+$. Assuming that radio-purity similar to Borexino is reached, a broader solar neutrino program is planned after the $^{130}$Te measurement. A major advantage for the SNO$^+$ measurement of $pep$ and CNO neutrinos is the 6000\,m.w.e of shielding provided by the rock overburden, reducing the $^{11}$C cosmogenic background by a factor of 100 with respect to Borexino.

Solar neutrinos have played a fundamental historical role in the discovery of the phenomenon of neutrino oscillations and thus non-zero neutrino mass. Even today, their study provides an important insight not only to solar and stellar physics, but also into the physics of the neutrino itself.  Solar neutrino physics is pinning down the LMA-MSW oscillation solution and is testing non-standard neutrino interaction models by measuring the energy-dependent electron neutrino survival probability $P_{ee}$ and matter effects. The latter can be tested by comparing the day and night rates of solar neutrinos, since only during the night the neutrinos cross the Earth. After the exclusion of a day-night asymmetry at 1\% precision level  at $^7$Be -neutrino energies performed by Borexino~\cite{BxDayNight}, Super-Kamiokande has successfully provided the first direct indication that neutrino oscillation probabilities are modified by the presence of matter by measuring the night-regeneration of electron flavour for higher-energy $^8$B solar neutrinos~\cite{MSWSuperK}. They determined the day-night asymmetry, defined as the difference of the average day and night rates, divided by their average, to be $(-3.2 \pm 1.1$ (stat) $\pm$ 0.5(syst))\%, which deviates from zero by 2.7$\sigma$. This asymmetry is consistent with neutrino oscillations for $4 \times 10^{-5} \rm{eV}^2 \leq \Delta$m$^2_{21} \leq 7 \times 10^{-5}$eV$^2$ and large mixing values of $\theta_{12}$, at the 68\% C.L. For what concerns $P_{ee}$, the vacuum-dominated low-energy region $< 1$ MeV is directly tested by Borexino data~\cite{LongPaper} while the matter-dominated high-energy region is tested at best by SNO~\cite{SNO} and Super-Kamiokande~\cite{SuperKPee} $^8$B data. It is also worth mention that the current best fit values of $\Delta m_{12}^2$ based on solar neutrino data and KamLAND reactor antineutrino data~\cite{gando} are in a slight tension at about 2$\sigma$ level~\cite{SuperKPee}.

\subsection{Supernova neutrinos}
\label{supernova}

  % Decoherence: E. Akhmedov
  % Detection: T. Mori.

A supernova (SN) explosion happens when a star at least 8 times more massive than the Sun collapses. Neutrinos with energies up to several tens of MeV carry away in a burst lasting few seconds almost 99\% of the released gravitational energy of the order of  $\sim$$3  \times 10^{53}$ erg~\cite{SNmodel}. Supernova neutrino detection would provide significant information about star collapse physics and also about neutrino physics, by studying the energy spectra, time profile, and flavour composition of SN neutrinos.

World-wide, several detectors currently running (IceCube, KamLAND, Super-Kamiokande, Baksan, LVD, Borexino, HALO) or nearing completion (SNO$^+$) are sensitive to a core-collapse supernovae neutrino signal in the Milky Way (expected  $\sim$3 SN/ 100 years). The neutrino signal emerges promptly from a supernova's core, whereas it may take hours for the first photons to be visible. Putting several detectors in coincidence can provide the astronomical community with a very high confidence early warning of the supernova's occurrence. For this purpose, an international SNEWS alarm system has been set-up~\cite{SNEWS} and soon will include also gravitational wave experiments.
The Super-Kamiokande 32\,kton water Cherekov detector would detect $\sim$7300 inverse beta decay interactions and $\sim$300 elastic scatterings from a 10 kpc distant SN, providing statistics sufficient to test SN models~\cite{Mori}. In spite of the multi-GeV energy threshold of Ice-Cube, the burst of galactic SN neutrinos would be recognized as a coincident increase in single PMT count rate. Liquid scintillator detectors expect about 700 events / kton, dominated by the inverse beta-decay and neutral-current $\nu$ + proton interactions.

The expected rate of SN from neighboring galaxy clusters (Mpc scale) is quite high, about 1~SN per year. However, for their detection, next generation multi-Mton detectors would be required.  

Neutrinos emitted from the past SN at Gpc scale are red-shifted by cosmic expansion and are often referred to as supernova diffuse or relic neutrinos (DSN or RSN). The expected flux of  $\bar \nu _e$~\cite{SRN} of about 20 cm$^{-2}$ s$^{-1}$ would lead to the detection of few  events per year per 20 kton (the size of the JUNO detector planned to start data-taking in 2020) in 10-30 MeV energy range. For their successful detection, an efficient technique to suppress the background from neutral current interaction of atmospheric neutrinos and fast neutrons is necessary. 

Super-Kamiokande's potential to detect RSN~\cite{SRN-SK} via inverse beta decay is limited mostly due to the low efficiency of detection of 2.2 MeV gamma\modf{s} from neutron capture on protons. The EGADS experiment is an ongoing R\&D program for the GAZOOKS! project~\cite{GADZOOKS}, in which Super-Kamiokande would be doped with Gadolinium. Gadolinium has a large cross section for thermal neutron capture, after which 3-4 $\gamma$'s of 8 MeV total energy are emitted and can be efficiently detected. 

The so called pre-supernova neutrinos are expected to be emitted during the Si-burning phase~\cite{preSN}. Their emission should start a few days before the core-collapse SN and their flux should gradually increase up to the SN burst. Due to the reduced flux, their detection would be possible only from relatively close stars, such as Antares (170 pc) or Betelgeuse (150 pc) and could be used as an early alarm of core-collapse burst. The study of their possible detection at KamLAND and GADZOOKS! is ongoing~\cite{Mori}.

In dense neutrino backgrounds present in supernovae and in the early Universe, neutrino oscillations may exhibit complex collective phenomena, such as synchronized oscillations, bipolar oscillations, and spectral splits and swaps. The effects of decoherence by wave packet separation on these phenomena have been studied for the first time by Akhmedov et al.~\cite{akhmedov}. It was shown that decoherence may modify the oscillation pattern significantly and can lead to qualitatively new effects, like for example strong flavour conversion, absent in the case of the usual MSW oscillations.

\subsection{Geoneutrinos}
\label{geonu}
  % S. Dye

Geoneutrinos, electron antineutrinos from natural $\beta$-decays of long-lived radioactive elements inside the Earth, bring to the surface unique information about our planet. The new techniques in neutrino detection opened a door into a completely new inter-disciplinary field at the border of particle physics and geology. The main goal of this study is to estimate the Earth's radiogenic heat, that is, the amount of heat released in the decays along the $^{238}$U and $^{232}$Th chains and that of $^{40}$K: a key parameter for both geophysical and geochemical models. The upper limit for the possible radiogenic heat is the measured integrated surface heat flux of $47 \pm 2$\,TW~\cite{davies}. The total radiogenic power contributing to this flux is not well known; different so-called Bulk Silicate Earth's models, summarized for example in~\cite{sramek}, predict a wide range of possible values in the range of $18 \pm 9$\,TW. While the crustal radiogenic heat is relatively well known to be $6.8^{+1.4}_{-1.1}$\,TW \cite{huang}, the mantle contribution is poorly known: its determination is the principal goal of geoneutrino studies.

The current experimental technique to detect geoneutrinos, using large-volume liquid scintillator detectors, is based on the inverse beta-decay reaction
%
%\begin{equation}
$\bar{\nu}_e + p \rightarrow e^+ + n$,
%\label{Eq:InvBeta}
%\end{equation}
%
enabling a space-time coincidence tag between the prompt (positron) and delayed ($\gamma$ from neutron capture) signal, strongly suppressing the background. The kinematic threshold for this reaction is 1.8 MeV and thus only a high-energy tail of  $^{238}$U and $^{232}$Th geoneutrinos can be detected, while $^{40}$K signal is completely out of reach.
In geoneutrino measurement\modf{s}, the most critical background sources are the antineutrinos from reactor power plants, while the dominant non-antineutrino backgrounds mimicking the coincidence tag are cosmogenically produced $^9$Li-$^8$He decaying ($\beta$ + n), ($\alpha$, n) interactions,  and accidental coincidences.

Only two underground experiments have succeeded in detecting geoneutrinos: Borexino in Italy and KamLAND in Japan, very far away from each other and placed in different geological environments. Borexino with its about 280 tons of liquid scintillator is about 3 times smaller with respect to KamLAND, but has much better radiopurity, so its non-antineutrino background is completely negligible. Both collaborations updated their results in 2013. Borexino has detected 14.3 $\pm$ 4.4 geoneutrinos in $(3.69 \pm 0.16) \times~10^{31}$ proton $\cdot$ year exposure collected from December 2007 to August 2012~\cite{bxgeo2}. KamLAND has detected $116^{+28}_{ -27}$ geoneutrino events in $(4.9 \pm 0.1) \times 10^{32}$ proton $\cdot$ year exposure collected from March 2002 and November 2012~\cite{gando}. A convenient unit for the measured signal is the normalized event rate, expressed as the so called Terrestrial Neutrino Unit (TNU), defined as the number of interactions detected during one year on a target of  $10^{32}$ protons ($\sim$1\,kton of liquid scintillator) and with 100\% detection efficiency. The Borexino and KamLAND results, expressed in this way, are $38.8 \pm 12$ and $ 30 \pm 7$ TNU, respectively.

The total abundance of the long-lived radioactive elements and the radiogenic heat are in a well fixed ratio, while the geoneutrino flux at the Earth's surface depends also on the distribution of the radioactive elements in the Earth interior. Thus, to extract from a measured geoneutrino flux the radiogenic heat is not straightforward and does include assumptions about the distribution of the heat producing elements. Both existent results are compatible with geological expectations, confirming that the new interdisciplinary field is born. However, mostly due to limited statistics, firm geologically significant results are not available yet. Keeping this in mind, the first indications of a measured mantle signal and the radiogenic heat do exist, as it was shown for example in~\cite{hindawi} combining both Borexino and KamLAND data. Both experiments will continue to take data, and some other experiments have geoneutrinos among their scientific goals (1 kton SNO$^+$ in Canada~\cite{Chen} starting soon and 20 kton JUNO in China planned for 2020).  A real breakthrough will come from the next generation of experiments, eventually placed in a carefully selected locations in which the crustal contribution to the signal can be well predicted~\cite{Jaupart} or is small, like for example on the oceanic crust (Hanohano project~\cite{hanohano}). Measuring geoneutrinos around the globe has also its importance because it can test the homogeneity of the Earth's interior.

\section{Summary}

Measurements of low energy neutrinos and antineutrinos have
  revealed the character of our natural world, and are expected to
  continue to produce interesting results.  Not only do measurements
  address the fundamental nature of these weakly-interacting
  particles, but they can also serve as unique probes of stars,
  supernovae, and the Earth.  The topics discussed during the
  ``Oscillation at low energies'' parallel session of the Neutrino
  Oscillation Workshop 2014 suggest that the coming years of neutrino
  research will continue to be fruitful.

%\section*{Acknowledgement}



%% References

\nocite{*}
%\bibliographystyle{elsarticle-num}
%\bibliography{NOW2014}



\end{document}